# Temperature dependent behavior of localized and delocalized electrons in nitrogen-doped 6H SiC crystals as studied by electron spin resonance

D. Savchenko,[1,2,a] E. Kalabukhova,[3] B. Shanina,[3] S. Cichoň,[1] J. Honolka,[1] V. Kiselov,[3] and E. Mokhov[4,5]

[1]*Institute of Physics AS CR, Prague, 182 21, Czech Republic*

[2]*National Technical University of Ukraine "Kyiv Polytechnic Institute", Kyiv, 03056, Ukraine*

[3]*V.E. Lashkaryov Institute of Semiconductor Physics, NAS of Ukraine, Kyiv, 03028, Ukraine*

[4]*A.F. Ioffe Physical Technical Institute, RAS, St. Petersburg, 194021, Russia*

[5]*Saint-Petersburg National Research University of Information Technologies, Mechanics and Optics, St. Petersburg, 19710, Russia*

We have studied the temperature behavior of the electron spin resonance (ESR) spectra of nitrogen (N) donors in n-type 6H SiC crystals grown by Lely and sublimation sandwich methods (SSM) with donor concentration of $10^{17}$ cm$^{-3}$ at $T$ = 60-150 K. A broad signal in the ESR spectrum was observed at $T \geq 80$ K with Lorentzian lineshape and $g_\parallel$ = 2.0043(3), $g_\perp$ = 2.0030(3), which was previously assigned in the literature to the N donors in the 1s(E) excited state. Based on the analysis of the ESR lineshape, linewidth and *g*-tensor we attribute this signal to the conduction electrons (CE). The emergence of the CE ESR signal at $T$ > 80 K was explained by the ionization of electrons from the 1s($A_1$) ground and 1s(E) excited states of N donors to the conduction band while the observed reduction of the hyperfine (hf) splitting for the $N_{k1,k2}$ donors with the temperature increase is attributed to the motion narrowing effect of the hf splitting. The temperature dependence of CE ESR linewidth is described by an exponential law (Orbach process) with the activation energy corresponding to the energy separation between 1s($A_1$) and 1s(E) energy levels for N residing at quasi-cubic sites ($N_{k1,k2}$). The theoretical analysis of the temperature dependence of microwave conductivity measured by the contact-free method shows that due to the different position of the Fermi level in two samples the ionization of free electrons occurs from the energy levels of $N_{k1,k2}$ donors in Lely grown samples and from the energy level of $N_h$ residing at hexagonal position in 6H SiC grown by SSM.

---

[a] Author to whom correspondence should be addressed. Electronic mail: dariyasavchenko@gmail.com.

## I. INTRODUCTION

A great amount of effort has been devoted to the investigation of nitrogen (N) donors in silicon carbide (SiC) polytypes by magnetic resonance methods.[1-9] However, the electron spin resonance (ESR) properties of N in n-type SiC with low and intermediate donor concentration ($N_D - N_A$) varying from $2 \cdot 10^{16}$ to $7 \cdot 10^{17}$ cm$^{-3}$ have so far been extensively studied at low temperatures when the donor electrons are bound in the ground state of the donor atoms. In this case, the ESR spectrum consists of two triplet lines due to the hyperfine (hf) interaction with $^{14}$N nuclei ($I = 1$, 99.6%) corresponding to N donors at quasi-cubic "k1" and "k2" sites ($N_{k1,k2}$), a line with a small unresolved hf splitting due to the N substituting hexagonal ("h") position ($N_h$) and a line triplet $N_x$ with $S = 1$ due to the distant donor pairs formed between the N atoms residing at quasi-cubic and hexagonal sites.[5,10]

Considerably less is known about the behavior of the N donors in the excited state. As was shown in Ref. [1-3] with the temperature increase the transition of the donor electrons from the symmetric 'hydrogen-like' 1s($A_1$) ground state to the thermally excited antisymmetric 1s(E) state (which has no hf interaction with the central $^{14}$N nuclei) occurs and is accompanied by the reduction of the hf splitting for $N_{k1}$, $N_{k2}$ and $N_h$ donors. This transition occurs at $T = 30 - 60$ K for $N_h$ donors[1] and at $T = 80 - 120$ K for $N_{k1,k2}$ donors[3], and the temperature interval can be shifted depending on the compensation degree of the sample. As it was suggested in Ref. [1, 3] the further increase of the temperature gives rise to the transformation of the triplet lines into one broad line, which was attributed to N in the thermally excited antisymmetric 1s(E) state.

In the present work we analyse the ESR lineshape, linewidth and $g$-tensor of the observed single broad line at $T \geq 80$ K in n-type 6H SiC wafers with intermediate N concentration of about ($N_D - N_A$) ≈ $10^{17}$ cm$^{-3}$ grown by Lely method and sublimation sandwich methods (SSM). From our results we conclude that this single line is caused by conduction electrons (CE) present in the n-type 6H SiC due to the ionization of the $N_{k1}$, $N_{k2}$, $N_h$ donor electrons from the ground $1s(A_1)$ and excited $1s(E)$ state to the conduction band. The contact-free microwave (MW) conductivity method has been employed to derive the energy characteristics of the two types of the samples. It was found that the ionization of free electrons to the conduction band occurs from the $N_{k1,k2}$ energy levels in Lely grown samples and from $N_h$ energy level in 6H SiC grown by SSM. We explain this fact by the different position of the Fermi level and different concentration of the shallow $N_h$ donors in the isolated electrically active state in two samples.

## II. MATERIALS AND METHODS

The n-type 6H SiC wafers with $(N_D - N_A) \approx (1-5) \cdot 10^{17}$ cm$^{-3}$ grown by modified Lely method and SSM[11-14] were investigated by x-ray photoelectron spectroscopy (XPS) and continuous wave ESR methods. The growth of the n-type 6H SiC wafers by modified Lely method was carried out around 2200-2400°C of the growth temperature and 30-50 mbar of Ar pressure with the growth rate of 1.2 mm/h on [0001] Si face using polycrystalline SiC as source materials. The growth of the n-type 6H SiC wafers by SSM was carried out at 1900$^0$C with the growth rate of 0.2 mm/h on the [0001]C face in a tantalum container under Si excess partial pressure using SiC micropowder with Si/C ~ 1.05 as a source of vapor composition. The size of the samples was about 7 x 4 x 0.3 mm.

The XPS measurements were performed in a NanoESCA spectrometer (Omicron Nanotechnology, Germany) using a monochromatic X-ray radiation source from an Al anode. The spectrometer also employs an energy-filtered PEEM (Photoemission Electron Microscope) with a Hg light source. The measured spot size on the samples was $100 \times 100$ μm$^2$. Prior to the insertion into the apparatus, the samples were immersed into a concentrated hydrogen fluoride water solution in an ultrasonic bath for 15 min, and then rinsed in pure water and dried under nitrogen flow. The samples were sputtered with 3 keV $Ar^+$ ions at a pressure of $7 \cdot 10^{-4}$ Pa for 90 min so that the bulk part of the material could be approached. The ion beam incidence angle was 15° from the normal to the sample surface. The XPS spectra were simulated with CasaXPS software using Gauss-Lorentzian peaks and Shirley backgrounds.

The ESR measurements were performed on a X-band (9.4 GHz) Bruker ELEXYS E580 spectrometer in the temperature range from 150 K to 60 K. The ESR experiments were carried out using the ER 4122 SHQE SuperX High-Q cavity using MW power level of 0.15 mW, a modulation frequency of 100 kHz, a modulation amplitude of 0.01-1 mT (depending on the ESR linewidth), and the conversion time of 120 ms. The ESR spectra simulation was performed using the EasySpin 5.0.1. software package.[15] The MW conductivity of the samples was estimated using contact-free method based on the change of the cavity $Q$-factor due to the absorption of the electrical E component of the MW field by free carriers.[16-21]

## III. EXPERIMENTAL RESULTS AND ANALYSIS

### A. The XPS spectra in Lely and SSM grown 6H SiC wafers

The XPS measurements revealed that the investigated samples are, in terms of chemical composition and surface morphology, spatially homogeneous. With respect to SSM grown samples, the higher doping level of the Lely-grown 6H-SiC samples resulted in a shift of the XPS C(1s) peaks to higher binding energies by ~0.4 eV. In both types of samples after the Ar sputtering the XPS C(1s) spectra (see Fig. 1) show an intense peak at ~283.5 eV caused by C-Si bonds.[22-25] In addition, a weak peak component at 284.8 eV was observed with the intensity about three times higher in the SSM grown 6H SiC sample than in the Lely grown one. The position of this component is typical for $sp^2$ graphite-like carbon.[22-25] The Si(2p) peaks measured in both types of the samples are composed of a single component at ~100.7 eV that corresponds to Si-C bonds.[22-25] Furthermore, the deviation from stoichiometry in the SSM grown 6H SiC samples was found to be not more than 10 %.

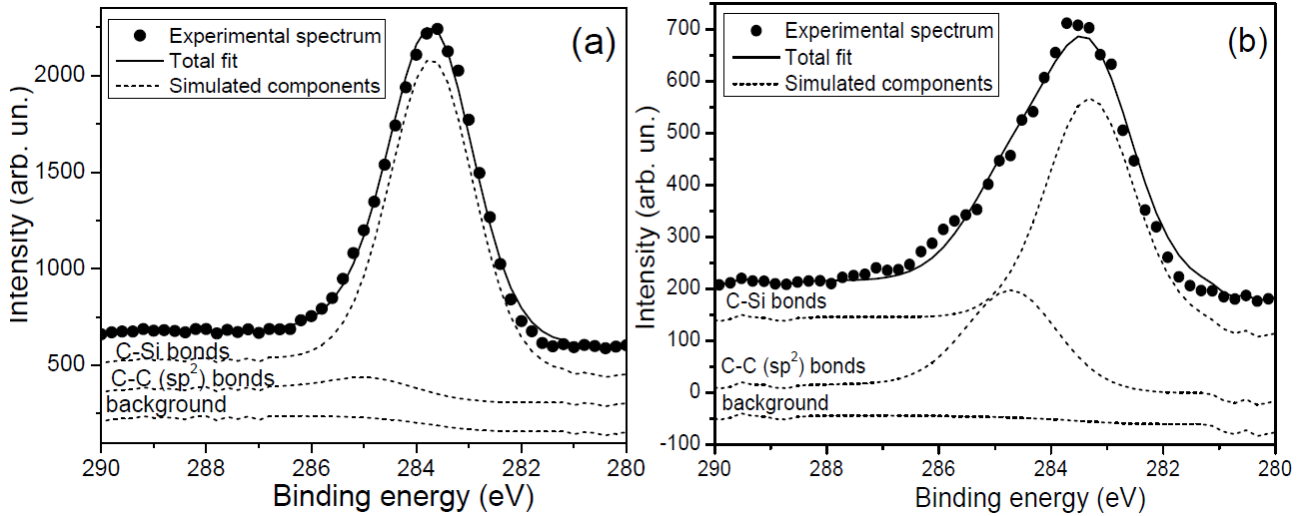

FIG. 1. The XPS spectra of the C(1s) peaks measured in Lely (a) and SSM (b) grown 6H SiC after Ar sputtering. Dots are experimental data, dashed lines are the simulated components and the solid lines are the sum of simulated components.

**B. The temperature behavior of the ESR spectra in Lely and SSM grown 6H SiC wafers**

Fig. 2 shows the experimental and simulated X-band ESR spectra of the N donors observed in Lely and SSM grown 6H SiC wafers at 60 K for $B \| c$. As it is seen from Fig. 2, the ESR spectra consist of two overlapping triplet lines from $N_{k1}$ ($g_\| = 2.0040(3)$, $g_\perp = 2.0026(3)$, $A_\| = A_\perp = 1.20$ mT) and $N_{k2}$ ($g_\| = 2.0037(3)$, $g_\perp = 2.0030(3)$, $A_\| = A_\perp = 1.19$ mT) donors.[5] The central line of $N_{k1,k2}$ triplets coincides with the ESR spectrum of $N_h$ donors ($g_\| = 2.0048(3)$, $g_\perp = 2.0028(3)$, $A_\| = 0.1$ mT, $A_\perp = 0.08$ mT).[1] Along with the ESR spectrum from the isolated N centers, the low intensity lines labeled $N_x$ ($g_\| = 2.0043(3)$, $g_\perp = 2.0029(3)$, $A_\| = A_\perp = 0.6$ mT)[5,10] are observed in between of the $N_{k1,k2}$ triplet lines. In accordance with Ref. [5], the $N_x$ lines are due to the triplet center with $S = 1$ responsible for the distant donor pairs between the N atoms residing at quasi-cubic and hexagonal sites.

From the simulation of the ESR spectra measured in the 6H SiC samples at 60 K (Fig. 2) the relative intensity ratio $I(N_{k2}) : I(N_{k1}) : I(N_h) : I(N_x)$ was found to be: 1.0 : 1.3 : 1.9 : 0.2 for the Lely grown and 1.0 : 0.9 : 3.0 : 0.1 for the SSM grown 6H SiC samples. Thus, the intensity of the ESR line from the $N_h$ donors with respect to that of $N_{k1,k2}$ triplet lines is significantly higher in the SSM grown 6H SiC than in the Lely grown 6H SiC sample, indicating that $N_h$ donors are mostly in the isolated state in 6H SiC grown by SSM. At the same time the intensity of the $N_x$ triplet is lower in the SSM grown sample than in the Lely grown 6H SiC.

The Fig. 3 shows ESR spectra of the N donors observed in Lely and SSM grown 6H SiC wafers in the temperature range from 60 K to 140 K. With the temperature increase, a decrease of the hf splitting constant for $N_{k1}$, $N_{k2}$ and $N_h$ donors was observed in both samples.

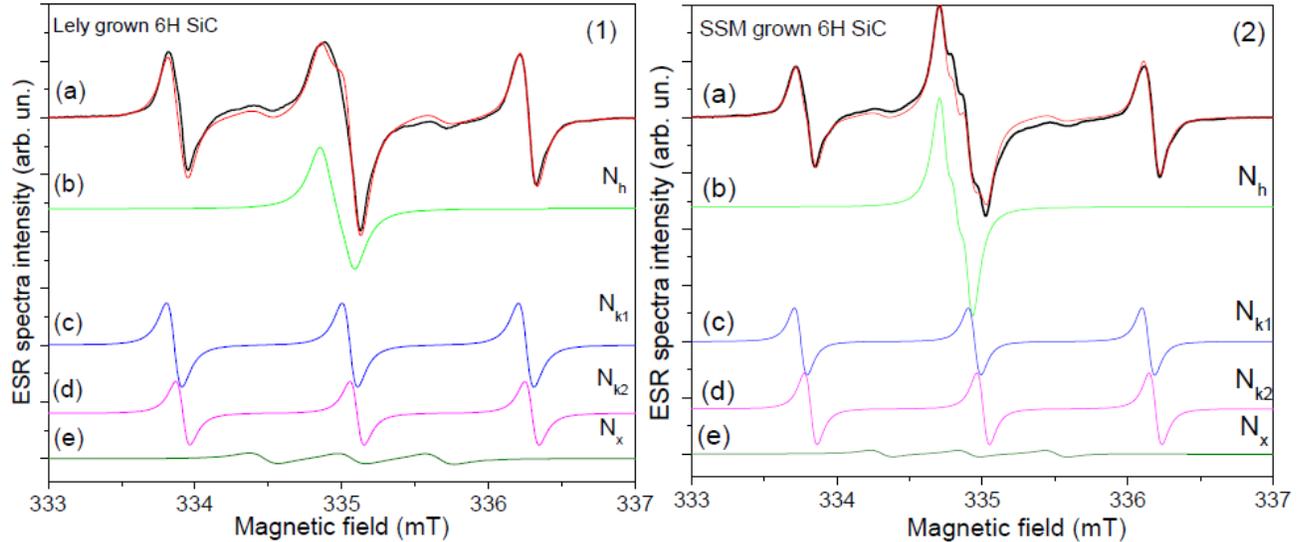

FIG. 2. The experimental (a, bold black line) ESR spectra measured in Lely-grown (1) and SSM-grown (2) 6H SiC wafers at $T = 60$ K along with the simulated components (b-e) and the sum of simulated components (a, thin red line). ***B*||*c***.

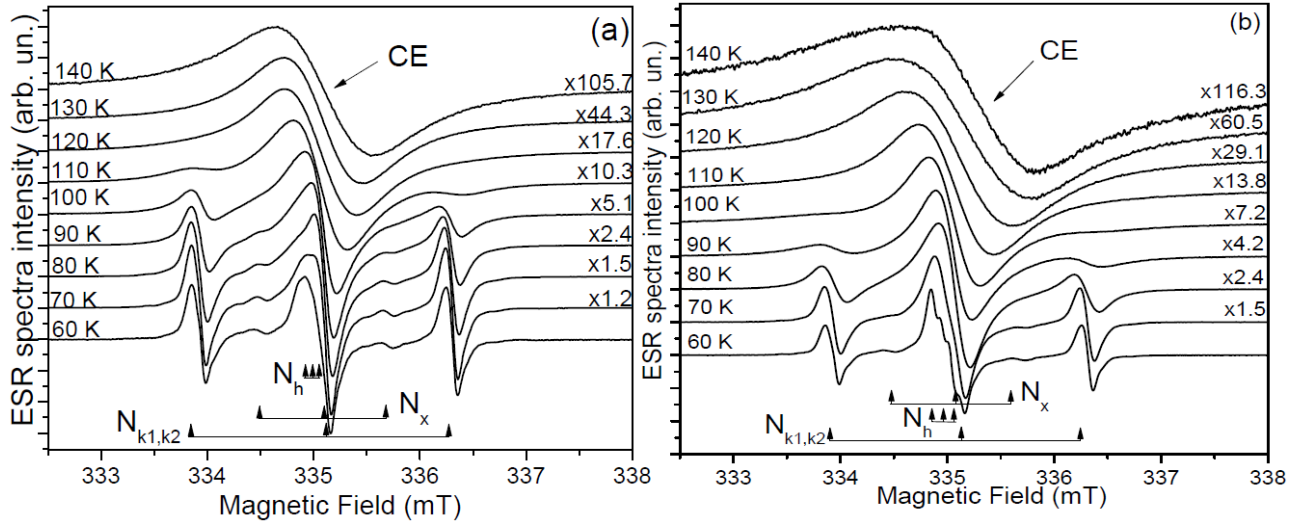

FIG. 3. The temperature behavior of the X-band ESR spectra measured in Lely-grown (a) and SSM-grown (b) 6H SiC wafers from 60 K to 140 K. *B*||*c*. On the right side of each spectrum the multiplication unit of the ESR spectra intensity is pointed with respect to the intensity of the ESR spectra measured at 60 K that was taken equal to 1.

The simulation components of the ESR spectra at different temperatures along with experimental ones are represented on Fig. 4a. Fig. 4b shows the experimental (scatters) and theoretically described (solid and dashed lines) temperature dependence of the hf splitting constant for $N_{k1,k2}$ donors. The reduction of the hf splitting for N donors was also observed previously[1-3] and qualitatively explained by the transition of the donor electrons from the symmetric 1s($A_1$) ground state to the antisymmetric excited 1s(E) state where the wave function would have no spin density at the nuclei site.

Following this model the reduction of the hf interaction constant *A* was caused by electron jumping from the ground 1s($A_1$) state with energy level $E_1$ to the excited 1s(E) state with energy level $E_2$. In this case, the average value of hf splitting constant *A* can be described by the sum $A = A_1 p_1 + A_2 p_2$, where the $p_1$, $p_2$ are the relative probabilities to find the electron in the 1s($A_1$) and 1s(E) states with hf splitting constants $A_1$, $A_2$, respectively and $A_2 \ll A_1$. The values $p_1$, $p_2$ are determined by the concentration of donors in two states and by their energy levels $E_1$ and $E_2$.

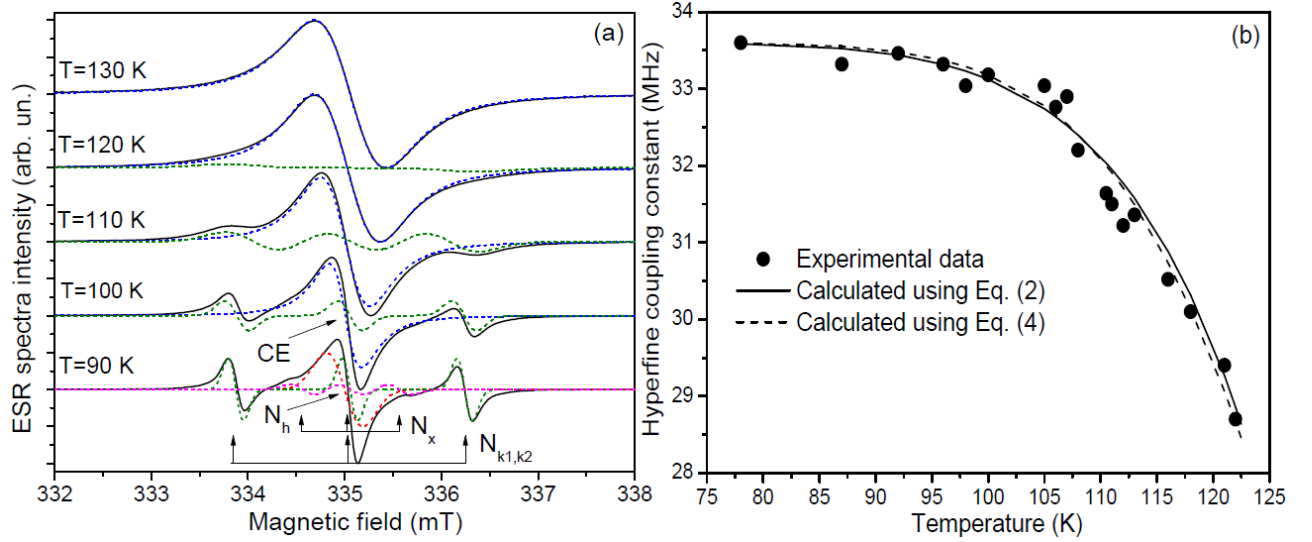

FIG. 4. The temperature evolution of the ESR spectra (a) and corresponding hf splitting of the $N_{k1,k2}$ donors (b) in Lely grown 6H SiC. The solid lines on Fig. 4a represent the experimental spectra and the dashed lines are the simulated components of the ESR spectra. The solid and dashed lines on Fig. 4b represent the result of fitting of Eq. (2) and Eq. (4) with experimental data, respectively.

As a result:

$$A(T) = A_1 \cdot \frac{n_1 \cdot \exp(-E_1/kT)}{n_1 \cdot \exp(-E_1/kT) + n_2 \cdot \exp(-E_2/kT)} + A_2 \cdot \frac{n_2 \cdot \exp(-E_2/kT)}{n_1 \cdot \exp(-E_1/kT) + n_2 \cdot \exp(-E_2/kT)}, \quad (1)$$

where $n_1$ and $n_2$ are the concentration of donors in $1s(A_1)$ and $1s(E)$ states. For the case when $E_2 > E_1$ and $A_2 \ll A_1$ the second term in Eq. (3) can be neglected. Hence:

$$A(T) = A_0 \frac{1}{1 + (n_2/n_1) \cdot \exp(-\Delta E/kT)}, \quad (2)$$

where $\Delta E = E_2 - E_1$. As was seen from Fig. 4b, the experimental curve is well described by the Eq. (2) (dashed line) with following parameters: $n_2/n_1 = 2.5 \cdot 10^4$; $\Delta E = 1450$ K = 124.3 meV. However, the obtained value of $\Delta E = 124.3$ meV does not correspond to the distance between the ground and excited state for N donors at the quasi-cubic sites amounting to 60.3 and 62.6 meV, for $N_{k1}$, $N_{k2}$, respectively[26] and therefore the reduction of the hf splitting cannot be explained by electron jumping from the ground $1s(A_1)$ to the excited $1s(E)$ state.

On the other hand the reduction of the hf splitting can be explained by the effect of motion narrowing. If the spins of donor electrons are in a rapid hopping motion, the local field corresponding to the hyperfine interaction with $^{14}N$ nuclei will fluctuate rapidly in time between $m_I = \pm 1$ and $m_I = 0$. In this case, the correlation function of the local field fluctuation is determined by the jumping frequency of the electron from one to another donor state and is determined by the temperature dependence of the correlation time $\tau_c$: $\tau_c(T) = \tau_{c0}\exp((k_B \cdot T_0/T)^k)$, where $k = 1, 1/2, 1/3, 1/4$, $\tau_{c0}$ – is the correlation time at high temperature, $k_B \cdot T_0$ is a barrier energy for an electron jump between the neighboring donors, $k_B$ – Boltzmann constant. Following A. Abragam,[27] the shape of the ESR signal with the hf splitting $\pm A$ and $I = 1$ can be written in the following form for the arbitrary $A \cdot \tau_c$ ratio (for the case $\tau_c^{-2} \ll A^2$):

$$I(\omega) = \frac{4A^2\tau_c^{-1}}{(\omega^2 - A^2)^2 + 4\omega^2\tau_c^{-2}} \approx \frac{2\tau_c^{-1}}{(\omega - A)^2 + \tau_c^{-1}} + \frac{2\tau_c^{-1}}{(\omega + A)^2 + \tau_c^{-1}} + \frac{2\tau_c^{-1}}{(\omega)^2 + \tau_c^{-1}} \qquad (3)$$

where $\omega = \omega_{res} - \omega_0$. Fig. 5 shows how the shape of the spectra changes for various values of $A\tau_c$. Likewise to the experimentally obtained temperature evolution of the $N_{k1,k2}$ ESR spectra shown in Fig. 4a the triplet lines calculated using Eq. (3) transform into one broad line with the increase of $A\tau_c$.

Equating the derivative of Eq. (3) to zero and finding the maximum of the function we obtain the following expression for $A$:

$$A(T) = A_0 \cdot \left(1 - \frac{2}{(A_0 \cdot \tau_c(T))^2}\right)^{1/2}, \qquad (4)$$

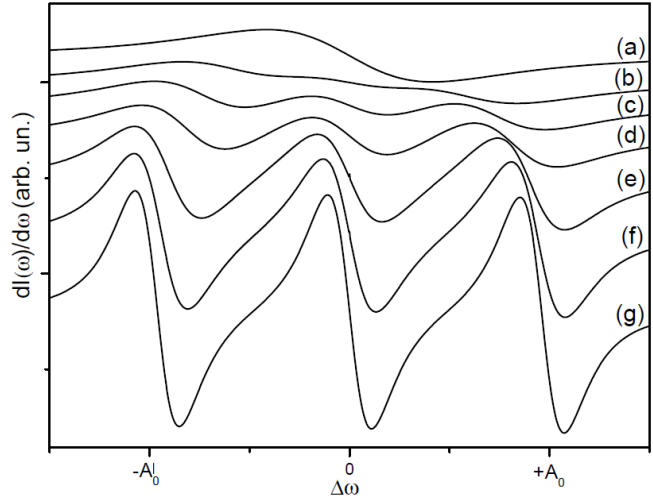

FIG. 5 The first derivative of theoretical spectra calculated using Eq. (4). The curves correspond to the following values of $A \cdot \tau_c$: a – 0.1, b – 1, c – 1.5, d – 2, e – 3, f – 4, g – 5.

where $A_0$ – is the hf splitting constant at low $T$.

As can be seen from Fig. 4b, the temperature evolution of the ESR spectrum of $N_{k1,k2}$ is described well by Eq. (2) (solid line) with $\tau_{c0} = 5.1 \cdot 10^{-4}$ s; and $k_B \cdot T_0 = 53.4$ meV. The obtained value of the 53.4 meV corresponding to the potential barrier of electron jumping from one to another donor site looks reasonable and gives us an argument to conclude that the reduction of the hf splitting for the $N_{k1,k2}$ donors is caused by the motional narrowing effect of the hf splitting.

Along with the decrease of hf splitting for $N_{k1,k2}$ centers a single line of a Lorentzian shape with $g_\parallel = 2.0043(3)$, $g_\perp = 2.0030(3)$ appears in the ESR spectrum of SSM grown 6H SiC at $T = 80$ K and in the ESR spectrum of Lely grown 6H SiC at $T = 100$ K (Figs. 3, 4a). Fig. 6 shows the temperature dependence of the linewidth $\Delta H_{pp}$ for this single line measured in two orientations of the magnetic field for both samples at $T > 80$ K, which can be well described by the following expression:

$$\Delta H_{pp}(T) = \Delta H_0 + c \cdot \exp(-\Delta / T) \qquad (5)$$

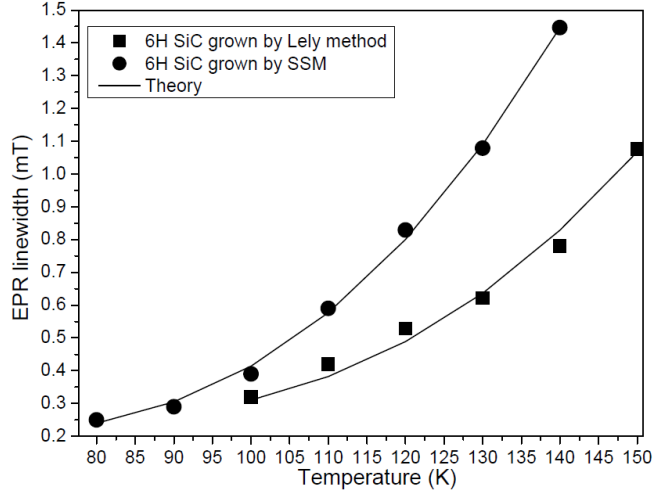

FIG. 6. The temperature behavior of the linewidth of the single ESR signal observed at $T > 80$ K in the 6H SiC crystals grown by Lely method and by SSM. $\mathbf{B}\|c$. The dots are experimental data, solid lines – the result of fitting of Eq. (5) with experimental data using the parameters given in Table 1.

where $\Delta H_0$ – is a temperature independent residual linewidth at $T = 0$.

The same exponential increase of the linewidth (Orbach process) was observed for CE in heavily doped 6H SiC above 40 K,[28] and was explained by the relaxation of the localized electrons (LE) via the excited levels with the energy $\Delta$ corresponding to the valley-orbit splitting ($\Delta E_{v.-o.}$) between the ground 1s($A_1$) and excited 1s(E) levels of N donors driven by the exchange interaction of CE with the LE, $c$ depends on the strength of the orbit-lattice coupling.

Hence, the similarity in the temperature behavior of the linewidth for CE in highly N doped 6H SiC and single line observed at $T > 80$ K in the ESR spectrum with $g_\| = 2.0043(3)$, $g_\perp = 2.0030(3)$ in 6H SiC crystals with intermediate N concentration allow us to attribute this line also to the CE. At the same time the CE ESR line integral intensity has weak temperature dependence at 80 K $< T <$ 150 K which is typical for the Pauli-like behavior.

The fact that the CE ESR signal has a Lorentzian lineshape is explained by the large skin depth, i.e. not so large conductivity. It is known that that for the samples with the size smaller than the skin depth, one expects a symmetric Lorentzian absorption spectrum for CE, whereas for the samples with larger size as compared with the skin depth an absorption and dispersion are of comparable results an asymmetric Dysonian line.[29,30]

The fitting parameters of Eq. (5) with the experimental data for the CE ESR linewidth obtained in n-type 6H SiC samples are given in Table 1. As can be seen from Table 1, the $\Delta$ value amounts to 700 K = 60 meV both for the Lely-grown and SSM-grown 6H SiC, which coincides with the value of the valley-orbit splitting ($\Delta E_{v.-o.}$) between the ground 1s($A_1$) and excited 1s(E) levels determined from the IR absorption spectra ($\Delta E_{v.-o.} = 60.3$ meV for $N_{k1}$ and $\Delta E_{v.-o} = 62.6$ meV for $N_{k2}$).[7]

Thus, the measurements of the CE linewidth temperature dependence can be considered as one of the ways to derive the values of the $\Delta E_{v.-o.}$ in SiC. The energy characteristics of the N donors in 6H SiC and 4H SiC can be also obtained via the measurements of the temperature dependence of the spin-relaxation time $T_1$ of the N donors at $T > 40$ K.[31] However, taking into account some discrepancies between the $\Delta E_{v.-o.}$ values obtained for the samples with different N concentration[31] and IR data,[7] this way looks less reliable.

TABLE 1. The $\Delta$, $\Delta H_0$, $c$ values obtained from the fitting of Eq. (5) with the experimental temperature dependence of CE ESR linewidth measured in Lely- and SSM-grown 6H SiC samples with $(N_D - N_A) \approx 10^{17}$ cm$^{-3}$ or $\boldsymbol{B} \| \boldsymbol{c}$.

| Sample | $\Delta$, K | $\Delta H_0$, mT | $c$, mT |
|---|---|---|---|
| Lely grown | 700 | 0.23 | 180 |
| SSM | 700 | 0.19 | 95 |

## C. The temperature dependence of the MW conductivity in Lely and SSM grown 6H SiC wafers

Considering that the CE play an important role not only in the magnetic properties of the N donors, but also in their electrical properties, we have studied the electrical characteristics of both n-type 6H SiC samples using the contact-free MW conductivity method based on the change of the ESR cavity $Q$-factor due to the absorption of the electrical E component of the MW field by the free carriers.[17-21]

Fig. 7 shows the temperature dependence of the cavity $Q$-factor loaded with the samples grown by Lely method and SSM along with an empty quartz sample holder in the temperature interval from 300 K to 50 K. A decrease of the cavity $Q$-factor with the temperature increase resulted from the variation of the MW conductivity was observed for both types of the 6H SiC wafers while the cavity $Q$-factor loaded with the empty sample holder remains unchanged in the whole temperature interval. As was shown in Ref. [28] in the highly compensated 6H SiC wafer (where the ESR spectra from N and boron are simultaneously observed) no temperature variation of $Q$-factor was observed as well. With the aim to derive the energy characteristics of the N donors from the temperature variation of $Q$-factor let us find the relationship between the variation of the $Q$-factor and the concentration of the free electrons $n_e$ in the samples.

It is known that the cavity $Q$-factor can be computed by summing the reciprocals of the unloaded $Q$-factor, denoted by $Q_u$, and loaded $Q$ due to the dielectric losses in the sample, denoted by $Q_e$:[32]

$$Q^{-1} = Q_u^{-1} + Q_e^{-1} \tag{6}$$

The overall $Q$ factor can be expressed as:

$$Q = \frac{Q_u}{1 + Q_u / Q_e}, \tag{7}$$

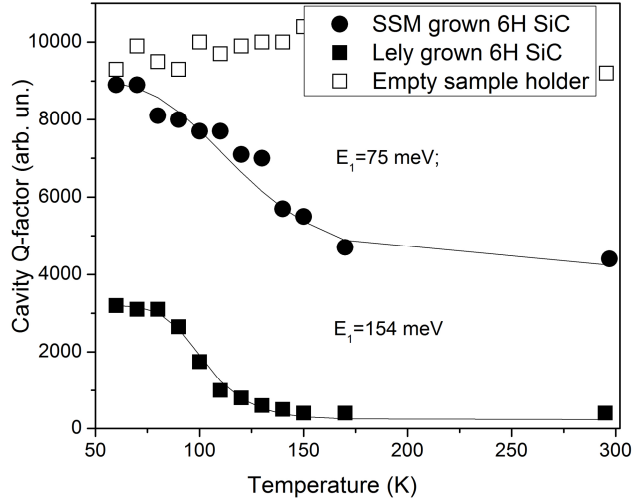

FIG. 7. The temperature behavior of the cavity $Q$-factor loaded with Lely-grown (solid squares) and SSM-grown (solid circles) 6H SiC samples and empty sample holder (open squares). Dots are experimental data, solid lines – the results of fitting of Eq. (12) with experimental data.

where $Q_e^{-1}$ is proportional to the imaginary part of the permittivity $\varepsilon''$, which is related to the energy lost in dielectric.[32,33]

On the other hand, within the frame of the Drude model for the electric conductor the $\varepsilon''$ has a contribution from the CE and is described by the expression:[33]

$$\varepsilon'' = \varepsilon_0'' + \frac{4\pi e^2 \lambda_0}{mc\gamma_c} \cdot n_e \qquad (8a)$$

Therefore:

$$Q_e^{-1} = Q_{e0}^{-1} + A n_e \qquad (8)$$

Substituting Eq. (8) into Eq. (7) one can obtain the expression:

$$Q = \frac{Q_u'}{1 + Q_u' \cdot A \cdot n_e}, \qquad (9)$$

where $Q_u' = \dfrac{Q_u}{1 + Q_u / Q_{0e}}$.

Thus, the variation of the $Q_e$ with the temperature is caused by a change of the concentration of the free electrons $n_e$ (sample conductivity). It turned out that the temperature dependence of $n_e$ is determined by Fermi statistics and charge neutrality equation. From the solution of the charge neutrality equation for the free electrons (CE) in the conduction band and the electrons remaining in the donor levels in condition of a weak compensation ($N_A < N_D/3$), it follows that the temperature function for the concentration of the electrons excited in the conduction band (free electrons $n_e$) can be written as:[34, 35]

$$n_e = \frac{2N_D}{1+\left(1+(8N_D/N_c)\cdot \exp((E_C-E_D)/kT)\right)^{1/2}}, \quad (10)$$

where $E_c$ – is the conduction band edge and parameter $N_c$ is the effective density of states in the conduction band:

$$N_c = 2.5\cdot 10^{19}\cdot (m^*/m)^{3/2}\cdot (T/300)^{3/2}, \quad (11)$$

with $m$ – denoting the free electron mass and $m^*$ is the effective mass of the electrons. The Eq. (10) is written for arbitrary ratio $R = (8N_D/N_c)$, because $R$ varies with the temperature increase from $R > 1$ to $R < 1$.

Substituting Eq. (10) into the Eq. (9) we can obtain the expression for the temperature dependence of the cavity $Q$-factor:

$$Q = \frac{C_1}{1+C_2\cdot n_e}, \quad (12)$$

where $C_1 = \dfrac{Q_u}{1+Q_u/Q_e}$ and $C_2 = C_1\cdot A$ are the numerical constant values.

The fit of the Eq. (12) (solid curve on Fig. 7) with the *Q*-factor experimental curves allows to determine the energy ionization $E_D$ of the N donors. Thus, for the Lely grown 6H SiC sample the energy ionization was found to be $E_D$ = 154 meV and for the SSM-grown sample: $E_D$ = 75 meV.

The obtained ionization energy for the Lely grown sample agrees well with that for $N_{k1}$ and $N_{k2}$ donors (137.6 meV and 142.4 meV),[36] while the activation energy determined for the sample grown by SSM coincides with that for the $N_h$ donors (81.04 meV).[36] This means that with temperature increase the $N_{k1k2}$ donors in Lely grown sample and $N_h$ donors in the sample grown by SSM are ionized and result in appearance of the free electrons (CE) in the conduction band.

Taking into account that the free carrier (electron) concentration $n_e$ is regulated by the position of the Fermi level with respect to the conduction band edge we can conclude that the position of the Fermi level in the Lely grown sample is close to the energy level of $N_{k1,k2}$, while in 6H SiC sample grown by SSM the Fermi level is fixed above the energy level of the $N_{k1,k2}$, indicating that the degree of compensation is higher in the Lely grown 6H SiC sample than that in 6H SiC grown by SSM. As a consequence, if the paramagnetic N donors are equally distributed between $N_{k1}$ and $N_{k2}$ energy levels the occupation of the $N_h$ donors is considerably higher in 6H SiC grown by SSM than that in Lely grown 6H SiC sample.

## IV. CONCLUSIONS

The temperature behavior of the ESR spectra of nitrogen (N) donors in n-type 6H SiC grown by the Lely method and SSM with N concentration of about $(N_D - N_A) \approx 10^{17}$ cm$^{-3}$ was studied in the temperature range at *T* = 4.2 – 150 K. The broad ESR signal with Lorentzian lineshape and $g_\parallel$ = 2.0043(3), $g_\perp$ = 2.0030(3) appeared in the ESR spectrum at *T* > 80 K was reassigned. Based on the analysis of the ESR lineshape, linewidth and *g*-tensor of the observed single broad line, we have attributed it to the CE appeared in the ESR spectrum of the n-type 6H SiC at high temperatures due to the ionization of the electrons from ground 1s(A) and excited 1s(E) state of the N donors to the conduction band.

We have found that the temperature dependence of CE ESR linewidth is described by the exponential law (Orbach process) in the temperature range from 80 K to 150 K with the activation energy corresponding to the energy separation between $1s(A_1)$ and $1s(E)$ energy levels for $N_{k1}$, $N_{k2}$ donors. The exponential increase of the CE ESR linewidth with the temperature is explained by coupling of the CE with the LE spin system.

The observed reduction of the hf splitting for the $N_{k1,k2}$ donors at $T > 75$ K with the temperature increase has been explained by electron jumping over the donor states, resulting in the fluctuation of local field (averaging of the hf interaction) determined by the temperature-dependent correlation time of fluctuations.

The electrical characteristics of both n-type 6H SiC samples were studied by the contact-free MW conductivity method. From the theoretical analysis of the temperature variation of the cavity $Q$-factor loaded with n-type 6H SiC samples we have found that the ionization of free electrons in the conduction band occurs from the $N_{k1,k2}$ energy levels in Lely grown sample and from $N_h$ energy level in 6H SiC sample grown by SSM. The ionization of free electrons from the different energy levels in two samples can be explained by the different position of the Fermi level and amounts of the distant donor pairs formed between N atoms residing at quasi-cubic and hexagonal sites in two samples. A small amount of the distant pairs in 6H SiC grown by SSM gives rise to the significantly higher concentration of the shallow $N_h$ donors in the isolated electrically active state and as a result leads to the ionization of the free electrons from $N_h$ energy level to the conduction band.

**ACKNOWLEDGMENTS**

The work was supported by GA ČR 13-06697P, SAFMAT CZ.2.16/3.1.00/22132 projects and Ministry of Education, Youth and Sports of Czech Republic Project No. LO1409. JH acknowledges the Purkyne fellowship program.